\documentclass[epjCONF]{svjour}
\usepackage{graphics}
\usepackage[varg]{txfonts} 
\usepackage[latin1]{inputenc}
\session-title{International Conference on New Frontiers in Physics(ICFP) 2012}
\begin{document}
\title{Clustering of Color Sources and the Equation of State of the QGP}
\author{Brijesh K Srivastava\thanks{\email{brijesh@purdue.edu}}}
\institute{Department of Physics, Purdue University, West Lafayette, IN-47906, USA}
\abstract{
Possible phase transition of strongly interacting matter from hadron to a quark-gluon plasma state have in the past received considerable interest. It has been suggested that this problem might be treated by percolation theory. The clustering of color sources with percolation (CSPM) is used to determine the equation of state (EOS) and the transport coefficient of the Quark-Gluon Plasma (QGP) produced in central A-A collisions at RHIC and LHC energies.                      
} 
\maketitle
\section{Introduction}
\label{intro}
One of the main goal of the study of relativistic heavy ion collisions is to study the deconfined matter, known as Quark-Gluon Plasma (QGP), which is expected to form at large densities. It has been suggested that the transition from hadronic to QGP state can be treated by the percolation theory \cite{celik}. The formulation of percolation problem is concerned with elementary geometrical objects placed on a random d-dimensional lattice. The objects have a well defined connectivity radius $\lambda$, and two objects can communicate if the distance between them is less than $\lambda$. Several objects can form a cluster of communication. At certain density of the objects a infinite cluster appears which spans the entire system. This is defined by the dimensionless percolation density parameter $\xi$ \cite{isich}.  Percolation theory has been applied to several areas ranging from clustering in spin system to the formation of galaxies. Figure 1 shows the transition from disconnected to connected system at high densities.
\begin{figure}
\centering        
\resizebox{0.70\textwidth}{!}{
\includegraphics{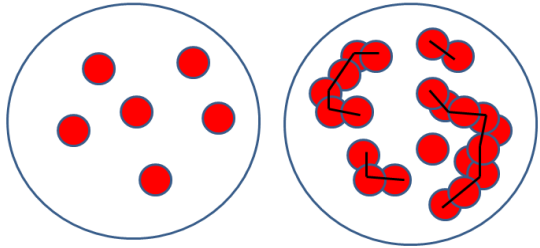}}
\caption{Left: Disconnected system. Right: Connected system} 
\label{perc1}
\end{figure}

 In nuclear collisions there is indeed, as a function of parton density, a sudden onset of large scale color connection. There is a critical density at which the elemental objects form one large cluster, loosing their independent existence. Percolation would correspond to the onset of color deconfinement and it may be a prerequisite for any subsequent QGP formation.

The determination of the EOS of hot, strongly interacting matter is one of the main challenges of strong interaction physics. Recent lattice QCD (LQCD) calculations for the bulk thermodynamic observables, e.g. pressure, energy density, entropy density and for the sound velocity have been reported \cite{bazavov}. 
 Recently, attention has been focused on the shear viscosity to entropy density ratio $\eta/s$  as a measure of the fluidity \cite{teaney,teaney1,lacey,rom}. The observed temperature averaged $\eta/s$, based on viscous hydrodynamics analyses of RHIC data, are suggestive of a strongly coupled plasma \cite{gul1,larry}.
In this talk a percolation model coupled with thermodynamical relations has been utilized to obtain EOS and the $\eta/s$ from the experimental data. 
\section{String Interactions and Percolation}
Multiparticle production at high energies is currently described in terms of color strings stretched between the projectile and target. Hardonizing these strings produce the observed hadrons. The strings act as color sources of particles through the creation of $q \bar{q}$ pairs from the sea. At low energies only valence quarks of nucleons form strings that then hadronize. The number of strings grows with the energy and with the number of nucleons of participating nuclei. Color strings may be viewed as small discs in the transverse space filled with the color field created by colliding partons. Particles are produced by the Schwinger mechanisms \cite{swinger}.  With growing energy and size of the colliding nuclei the number of strings grow and start to overlap to form clusters \cite{pajares1,pajares2}. At a critical density a macroscopic cluster appears that marks the percolation phase transition. 2D percolation is a non-thermal second order phase transition,
but in CSPM the Schwinger barrier penetration mechanism for particle production and the fluctuations in the associated string tension due to the strong string interactions make it possible to define a temperature.
Consequently the particle spectrum is "born" with a thermal distribution \cite{bialas}. 
With an increasing number of strings there is a progression from isolated individual strings to clusters and then to a large cluster which suddenly spans the area. In two dimensional percolation theory the relevant quantity is the dimensionless percolation density parameter given by \cite{pajares1,pajares2}  
\begin{equation}  
\xi = \frac {N S_{1}}{S_{N}}
\end{equation}
where N is the number of strings formed in the collisions and $S_{1}$ is the transverse area of a single string and $S_{N}$ is the transverse nuclear overlap area. The critical cluster which spans $S_{N}$, appears for
$\xi_{c} \ge$ 1.2 \cite{satz1}. As $\xi$ increases the fraction of $S_{N}$ covered by this spanning cluster increases.

The percolation theory governs the geometrical pattern of string clustering. It requires some dynamics to describe the interaction of several overlapping strings. It is assumed that a cluster behaves as a single string with a higher color field corresponding to the vectorial sum of the color charge of each individual string. Knowing the color charge, one can calculate the multiplicity $\mu_{n}$ and the mean transverse momentum squared $\langle p_{t}^{2} \rangle$ of the particles produced by a cluster of strings. One finds \cite{pajares1,pajares2}  
\begin{equation}  
\mu_{n}= \sqrt {\frac {nS_{n}}{S_{1}}}\mu_{1}
\end{equation}
\begin{equation}  
\langle p_{t}^{2} \rangle_{n}= \sqrt {\frac {nS_{1}}{S_{n}}\langle p_{t}^{2} \rangle_{1}}
\end{equation}
where $\mu_{1}$ and $\langle p_{t}^{2}\rangle_{1}$ are the mean multiplicity and average transverse momentum squared of particles produced by a single string . In the saturation limit, all the strings overlap into a single cluster that approximately occupies the whole interaction area, one gets the following universal scaling law 

\begin{equation}                                      
\langle p_{t}^{2}\rangle_{n}  = \frac {S_{1}}{S_{n}}\frac {\langle p_{t}^{2}\rangle_{1}}{\mu_{n}}{\mu_{1}}    
\end{equation} 
 
In the limit of high density one obtains
\begin{equation}
\langle \frac {nS_{1}}{S_{n}} \rangle = 1/F^{2}(\xi)
\end{equation}
with
\begin{equation}
F(\xi) = \sqrt {\frac {1-e^{-\xi}}{\xi}}
\end{equation}
being the color suppression factor. It follows that 
\begin{equation}
\mu = N F(\xi)\mu_{1},  \langle p_{t}^{2}\rangle = \frac {1}{F(\xi)}\langle p_{t}^{2}\rangle_{1}
\end{equation}
A similar scaling is found in the Color Glass Condensate approach (CGC)\cite{cgc,perx}. The saturation scale $Q_{s}$ in CGC corresponds to $ {\langle p_{t}^{2} \rangle_{1}}/F(\xi)$ in CSPM.
The net effect due to $F(\xi)$ is the reduction in hadron multiplicity and increase in the average transverse momentum of particles. The CSPM model calculation for hadron multiplicities and momentum spectra was found to be in excellent agreement with experiment \cite{diasde,diasde2}.

\section{Color Suppression Factor  $F(\xi)$}
The suppression factor is determined by comparing the $\it pp$ and  A+A transverse momentum spectra. 
To evaluate the initial value of $F(\xi)$ from data for Au+Au collisions, a parameterization of $\it pp$ events at 200 GeV  is used to compute the $p_{t}$ distribution \cite{nucleo,levente,eos}
\begin{equation}
dN_{c}/dp_{t}^{2} = a/(p_{0}+p_{t})^{\alpha}
\end{equation}
where a is the normalization factor.  $p_{0}$ and $\alpha$ are parameters used to fit the data. This parameterization  also can be used for nucleus-nucleus collisions to take into account the interactions of the strings \cite{pajares2}
\begin{equation}
dN_{c}/dp_{t}^{2} = \frac {a'}{{(p_{0}{\sqrt {F(\xi_{pp})/F(\xi)}}+p_{t})}^{\alpha}}
\end{equation}
In pp collisions $F(\xi) \sim$ 1 at these energies due to the low overlap probability.
Figure 2 shows a plot of $F(\xi)$ as a function of charged particle multiplicity per unit transverse area $\frac {dN_{c}}{d\eta}/S_{N}$ for Au+Au collisions at 200 GeV for various centralities for the STAR data \cite{nucleo,levente,eos}.  
$F(\xi)$ decreases in going from peripheral to central collisions. The $\xi$ value is obtained using Eq. (6), which increases with the increase in centrality. These experimental values of $F(\xi)$ are used to obtain temperature, EOS and $\eta/s$.  

\begin{figure}[thbp]
\centering        
\vspace*{-0.5cm}
\resizebox{0.55\textwidth}{!}{
\includegraphics{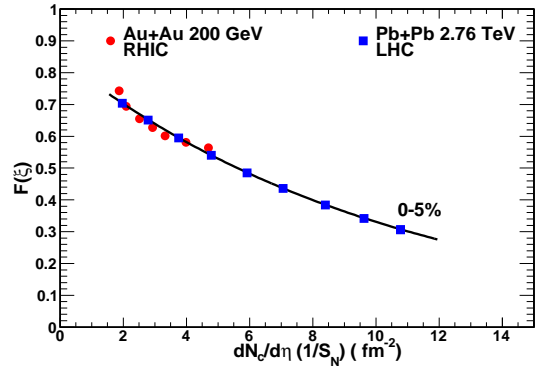}
}
\vspace*{-0.1cm}
\caption{ Color suppression factor $F(\xi)$ as a function of $\frac {dN_{c}}{d\eta}/S_{N}(fm^{-2})$. 
The solid red circles are for Au+Au collisions at 200 GeV(STAR data) \cite{nucleo}. The error is smaller than the size of the symbol. The line is fit to the STAR data. The solid blue squares are for  Pb+Pb at 2.76 TeV.} 
\end{figure}  

\section{Temperature}
The connection between the measured $\xi$ and the temperature $T(\xi)$ involves the Schwinger mechanism (SM) for particle production. 
The Schwinger distribution for massless particles is expressed in terms of $p_{t}^{2}$ \cite{swinger,wong}
\begin{equation}
dn/d{p_{t}^{2}} \sim e^{-\pi p_{t}^{2}/x^{2}}
\end{equation}
where the average value of the string tension is  $\langle x^{2} \rangle$. The tension of the macroscopic cluster fluctuates around its mean value because the chromo-electric field is not constant.
The origin of the string fluctuation is related to the stochastic picture of 
the QCD vacuum. Since the average value of the color field strength must 
vanish, it can not be constant but changes randomly from point to point \cite{bialas}. Such fluctuations lead to a Gaussian distribution of the string tension for the cluster, which gives rise to the thermal distribution \cite{bialas,eos}
\begin{equation}
dn/d{p_{t}^{2}} \sim e^{(-p_{t} \sqrt {\frac {2\pi}{\langle x^{2} \rangle}} )}
\end{equation}
with $\langle x^{2} \rangle$ = $\pi \langle p_{t}^{2} \rangle_{1}/F(\xi)$. 
The temperature is expressed as  
\begin{equation}
T(\xi) =  {\sqrt {\frac {\langle p_{t}^{2}\rangle_{1}}{ 2 F(\xi)}}}
\end{equation} 
\section{Equation of State}
Among the most important and fundamental problems in finite-temperature QCD are the calculation of the bulk properties of hot QCD matter and characterization of the nature of the QCD phase transition. 
The QGP according to CSPM is born in local thermal equilibrium  because the temperature is determined at the string level. We use CSPM and thermodynamical relations  to calculate energy density, entropy density and sound velocity (EOS).  After the initial temperature $ T > T_{c}$ the  CSPM perfect fluid may expand according to Bjorken boost invariant 1D hydrodynamics \cite{bjorken}
\begin{eqnarray}
\frac {1}{T} \frac {dT}{d\tau} = - C_{s}^{2}/\tau  \\
\frac {dT}{d\tau} = \frac {dT}{d\varepsilon} \frac {d\varepsilon}{d\tau} \\
\frac {d\varepsilon}{d\tau} = -T s/\tau \\
s =(1+C_{s}^{2})\frac{\varepsilon}{T}\\
\frac {dT}{d\varepsilon} s = C_{s}^{2} 
\end{eqnarray}
where $\varepsilon$ is the energy density, s the entropy density, $\tau$ the proper time, and $C_{s}$ the sound velocity. Above the critical temperature only massless particles are present in CSPM. 
The initial energy density $\varepsilon_{i}$ above $T_{c}$ is given by \cite{bjorken}
\begin{equation}
\varepsilon_{i}= \frac {3}{2}\frac { {\frac {dN_{c}}{dy}}\langle m_{t}\rangle}{S_{n} \tau_{pro}}
\end{equation}
To evaluate $\varepsilon_{i}$ we use the charged pion multiplicity $dN_{c}/{dy}$ at midrapidity and $S_{n}$ values from STAR for 0-10\% central Au+Au collisions at $\sqrt{s_{NN}}=$200 GeV \cite{levente}. The factor 3/2 in Eq.(17) accounts for the neutral pions. The average transverse mass $\langle m_{t}\rangle$ is given by $\langle m_{t}\rangle = \sqrt {(\langle p_{t}\rangle^2 + m_{0}^2)}$, where $\langle p_{t}\rangle$ is the transverse momentum of pion and $m_{0}$ being the mass of pion.
The dynamics of massless particle production has been studied in two-dimensional quantum electrodynamics (QED2).
QED2 can be scaled from electrodynamics to quantum chromodynamics using the ratio of the coupling constants \cite{wong}. The production time $\tau_{pro}$ for a boson (gluon) is \cite{swinger}.  
\begin{equation}
\tau_{pro} = \frac {2.405\hbar}{\langle m_{t}\rangle}
\end{equation}
From the measured value of  $\xi$ and $\varepsilon$ 
it is found  that $\varepsilon$ is proportional to $\xi$ for the range 
$1.2 < \xi < 2.88$.  Figure 3 shows a plot of energy density as a function of $\xi$. $\varepsilon_{i}= 0.788$ $\xi$ GeV/$fm^{3}$ \cite{nucleo,levente}.

\begin{figure}[thbp]
\centering        
\vspace*{-0.2cm}
\resizebox{0.55\textwidth}{!}{
  \includegraphics{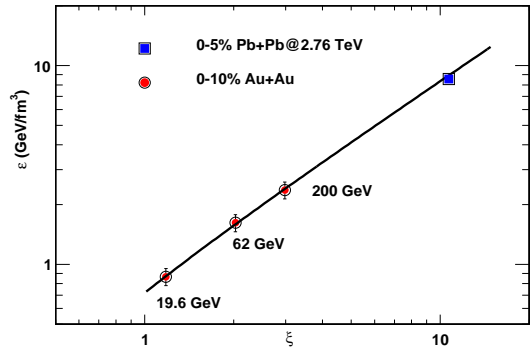}
}
\vspace*{-0.2cm}
\caption{Energy density $\epsilon$ as a function of the percolation density parameter $\xi$. The value for LHC energy is shown as blue square.} 
\end{figure}
 Figure 4 shows a plot of $\varepsilon/T^{4}$  as a function of T/$T_{c}$. The lattice QCD results are from HotQCD Collaboration \cite{bazavov}.  CSPM is in excellent agreement with the LQCD calculations in the phase transition region for $T/T_{c} \leq $1.5. The sound velocity and entropy density results are also in good agreement with LQCD calculations \cite{eos}.

\begin{figure}[thbp]
\centering        
\vspace*{-0.2cm}
\resizebox{0.55\textwidth}{!}{
\includegraphics{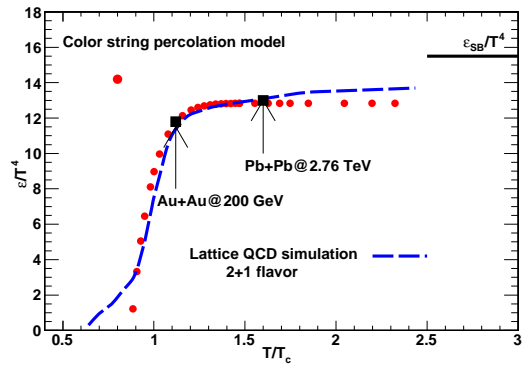}
}
\vspace*{-0.2cm}
\caption{ $\varepsilon/T^{4}$ as a function of T/$T_{c}$.The lattice QCD calculation is shown as dotted blue line \cite{bazavov}. } 
\end{figure}     
\section{Shear Viscosity}
The relativistic kinetic theory relation for the shear viscosity over entropy density ratio, $\eta/s$ is given by \cite{gul1,gul2}
\begin{equation}
\frac {\eta}{s} \simeq \frac {T \lambda_{mfp}}{5}     
\end{equation}
where T is the temperature and $\lambda_{mfp}$ is the mean free path
$\lambda_{mfp} \sim \frac {1}{(n\sigma_{tr})}$.
$\it n $ is the number density of an ideal gas of quarks and gluons and $\sigma_{tr}$ the transport cross section for these constituents. In CSPM the number density is given by the effective number of sources per unit volume 
\begin{equation}
n = \frac {N_{sources}}{S_{N}L}
\end{equation}
 L is the longitudinal extension of the source, L = 1 $\it  fm $. The effective no. of sources is given by the total area occupied by the strings divided by the effective area of the string $S_{1}F(\xi) $ \cite{eos2}. 
\begin{equation}
N_{sources} = \frac {(1-e^{-\xi}) S_{N}}{S_{1} F(\xi)} 
\end{equation}
$\eta/s$ is obtained from $\xi$ and the temperature
\begin{equation}
\frac {\eta}{s} ={\frac {TL}{5(1-e^{-\xi})}} 
\end{equation}
 Figure 5 shows a plot of $\eta/s$ as a function of T/$T_{c}$. 
The lower bound shown in Fig. 5 is given by AdS/CFT \cite{kss}. 
The theoretical estimates of $\eta/s$ obtained for both the weakly (wQGP) and strongly (sQGP) coupled QCD plasma are shown in Fig. 5 \cite{gul1}. It is seen that at the RHIC top energy  $\eta/s$ is close to the sQGP. Even at the LHC energy it follows the trend of the sQGP. By extrapolating the $\eta/s$ CSPM values to higher temperatures it is clear that $\eta/s$ could approach the weak coupling limit near $T/T_{c}$ $\sim$ 5.8. 
\begin{figure}[thbp]
\centering        
\vspace*{-0.2cm}
\resizebox{0.55\textwidth}{!}{
\includegraphics{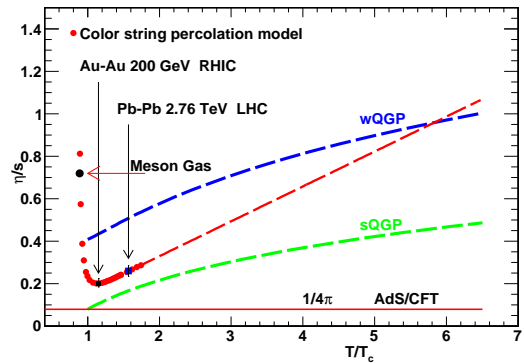}
}
\vspace*{-0.1cm}
\caption{$\eta/s$ as a function of T/$T_{c}$. Au+Au at 200 GeV for 0-10$\%$ centrality is shown as solid black square. wQGP and sQGP values are shown as dotted blue and green lines respectively \cite{gul1}. The estimated value for Pb+Pb at 2.76 TeV for 0-5$\%$ centrality  is shown as a solid blue square. The red dotted line represents the extrapolation to higher temperatures from the CSPM. The hadron gas value for $\eta/s$ $\sim$ 0.7 is shown as solid black circle at T/$T_{c} \sim $0.88 \cite{meson}.} 
\end{figure} 
\section{RHIC to LHC} 
 The STAR results for Au+Au collisions at $\sqrt{s_{NN}}$ = 200 GeV have been extrapolated to estimate F$(\xi)$ values for Pb+Pb collisions at $\sqrt{s_{NN}}$=2.76 TeV. The STAR points in Fig.2 has the functional form 
\begin{equation}
F(\xi) = exp [-0.165 -0.094 \frac {dN_{c}}{d\eta}/S_{N}]
\end{equation}
Recently, the ALICE experiment at LHC published the charged-particle multiplicity density data as a function of centrality in Pb+Pb collisions at $\sqrt{s_{NN}}$ = 2.76 TeV \cite{alice1}. The ALICE data points are shown in Fig.2. For central 0-5$\%$ in Pb+Pb collisions  $\xi$ = 10.56 as compared to $\xi$ = 2.88 for central Au+Au collisions at 200 GeV. The extrapolated value of $\varepsilon$ for central Pb+Pb collision at 2.76 TeV is  8.32 $GeV/fm^3$ as  shown in Fig.3. For Pb+Pb collisions the temperature is $\sim$ 262.2 MeV for 0-5$\%$ centrality, which is $\sim$ 35 $\%$ higher than the temperature from  Au+Au collisions \cite{eos}.

One way to verify the validity of extrapolation from RHIC to LHC energy is to compare the energy density expressed as $\varepsilon/T^4$ with the available lattice QCD results. The LHC point is shown in  Fig. 4. It is observed that at LHC energy the CSPM results are in excellent agreement with the lattice QCD results. The lattice and CSPM results are available for T/$T_{c} < 2$.

The estimated value of $\eta/s$ for Pb+Pb is shown in Fig. 5 at T/$T_{c}$ = 1.57. These results from STAR and ALICE data show that the $\eta/s$ value is 2.5 and 3.3 times the KSS bound \cite{kss}. 
\section{Summary}
Two central objectives in the experimental characterization of the QGP are the Equation Of State (EOS) and the shear viscosity to entropy ratio $\eta/s$.
The percolation analysis of the color sources applied to STAR data at RHIC provides a compelling argument that the QGP is formed in central Au+Au collisions at $\sqrt{s_{NN}}=$ 200 GeV. It also suggests that the QGP is produced in all soft high energy high multiplicity collisions when the string density exceeds the percolation transition. We found $\eta/s$ = 0.204 $\pm 0.020$ at $T/T_{c}$ = 1.15 ( RHIC ) and $\eta/s$ =0.260 $\pm 0.020$ at $T/T_{c}$ = 1.57 (LHC). 
In the phase transition region $\eta/s$ is 2-3 times the conjectured quantum limit for RHIC to LHC energies. 
The whole picture is consistent with the formation of a fluid with a low shear to viscosity ratio.
 Thus Clustering and percolation can provide a conceptual basis for the QCD phase diagram which is more general than the symmetry breaking \cite{satzx}.
\section{Acknowledgement}     
 This research was supported by the Office of Nuclear Physics within the U.S. Department of Energy  Office of Science under Grant No. DE-FG02-88ER40412.

\end{document}